\documentclass[10pt]{article}
\usepackage{amsmath}
\usepackage{amssymb}
\usepackage{graphicx}

\usepackage{bm}
\def\dd{\mathrm{d}}
\def\GR{\mathrm{GR}}

\begin{document}

\setcounter{figure}{0}
\setcounter{table}{0}
\setcounter{footnote}{0}
\setcounter{equation}{0}

\vspace*{0.5cm}

\noindent {\Large TESTING GRAVITY LAW IN THE SOLAR SYSTEM}
\vspace*{0.7cm}

\noindent\hspace*{1.5cm} B. LAMINE$^1$, J.-M. COURTY$^1$,
S. REYNAUD$^1$, M.-T. JAEKEL$^2$  \\
\noindent\hspace*{1.5cm} $^1$ Laboratoire Kastler Brossel\\
\noindent\hspace*{1.5cm} CNRS : UMR8552 - UPMC Univ Paris 06 - \'Ecole
Normale Sup\'erieure de Paris - ENS Paris
	Case 74 - 4 place Jussieu, F-75252 Paris CEDEX 05\\
\noindent\hspace*{1.5cm} e-mail: brahim.lamine@upmc.fr,
jean-michel.courty@upmc.fr and serge.reynaud@upmc.fr\\
\noindent\hspace*{1.5cm} $^2$ Laboratoire de Physique Th\'eorique\\
\noindent\hspace*{1.5cm} CNRS : UMR8549 - UPMC Univ Paris 06 - \'Ecole Normale Sup\'erieure de Paris - ENS Paris
	24 rue Lhomond, 75231 Paris CEDEX 05\\
\noindent\hspace*{1.5cm} e-mail: jaekel@lpt.ens.fr\\

\vspace*{0.5cm}

\noindent {\large ABSTRACT.}
The predictions of General relativity (GR) are in good agreement with
observations in the solar system. Nevertheless, unexpected anomalies
appeared during the last decades, along with the increasing precision
of measurements. Those anomalies are present in spacecraft tracking
data (Pioneer and flyby anomalies) as well as ephemerides. In
addition, the whole theory is challenged at galactic and cosmic scales
with the dark matter and dark energy issues. Finally, the unification
in the framework of quantum field theories remains an open question,
whose solution will certainly lead to modifications of the theory,
even at large distances. As long as those ``dark sides'' of the
universe have no universally accepted interpretation nor are they
observed through other means than the gravitational anomalies they
have been designed to cure, these anomalies may as
well be interpreted as deviations from GR. In this context, there is a
strong motivation for improved and more systematic tests of GR inside
the solar system, with the aim to bridge the gap between gravity
experiments in the solar system and observations at much larger
scales. We review a family of metric extensions of GR which preserve
the equivalence principle but modify the coupling between energy and
curvature and provide a phenomenological framework which generalizes the PPN
framework and ``fifth force'' extensions of GR. We briefly discuss
some possible observational consequences in relation with highly
accurate ephemerides.

\vspace*{1cm}

\noindent {\large 1. TESTS OF GENERAL RELATIVITY (GR) IN THE SOLAR SYSTEM}
\smallskip

The foundations of GR rely on two main pillars. The first one is the
equivalence principle which states the universality of free fall and
gives gravitation its geometric nature. This principle is tested in
modern experiments at the $10^{-12}$ level, which makes it one of the
best tested properties of nature. The validity of the equivalence
principle has also been tested very accurately in the solar system
using Lunar Laser Ranging [1] or the Sun-Mars orbit [2]. The resulting
bound is too small to allow the comparatively large anomalies observed
in the solar system. Therefore, even if a violation of the equivalence
principle is nevertheless possible, this strongly indicates that the
anomalies, if of gravitational origin, should find an explanation in
the framework of metric extensions of GR. We focuss on theories which
describe gravity by a tensor metric field $g_{\mu\nu}$. Let us mention
at this point that the general static and isotropic metric, which can
be used as a preliminary description of the solar system, essentially
reduces to two functions $g_{00}(r)$ and $g_{rr}(r)$ such that
\begin{equation}
  \label{eq:8}
  \dd s^2\equiv g_{\mu\nu}\dd x^\mu\dd x^\nu=g_{00}(c\dd
  t)^2+g_{rr}\bm{\dd r}^2
\end{equation}

\noindent The Einstein curvature tensor $G_{\mu\nu}$, built
upon the Ricci curvature, has a null covariant divergence (Bianchi identity)
\begin{equation}
  \label{eq:2}
  G_{\mu\nu}\equiv R_{\mu\nu}-\frac{1}{2}\,g_{\mu\nu}R\qquad,\qquad
  D^\mu G_{\mu\nu}=0
\end{equation}

This geometrical identity is often put in correlation with the
conservation of the stress tensor, $D^\mu T_{\mu\nu}=0$. This remark
leads to the second pillar of GR which are the equations relating the
curvature of spacetime to the energy-momentum content. Those equations
allow one to determine the metric tensor from the distribution of
energy-momentum in spacetime through the Einstein-Hilbert equation
which involves a unique Newton gravitational constant $G_N$
\begin{equation}
   G_{\mu\nu}=\frac{8\pi G_N}{c^4}T_{\mu\nu}\label{eq:EI}
\end{equation}

Note that this form is not imposed by any geometrical argument, so
that GR has to be selected out from a large variety of metric theories
by comparing the predictions drawn from the Einstein-Hilbert equation
to the results of observations or experiments. When performed in the
solar system, the tests effectively show a good agreement with the
solutions of (\ref{eq:EI}), which means that the metric tensor
$g_{\mu\nu}$ has a form close to its GR prediction given, for a static
point-like mass $M$ by
\begin{equation}
g_{00}=1+2\phi+2\phi^2+\ldots\quad,\quad-g_{rr}=1-2\phi+\ldots\quad
 \text{with}\quad
\phi=-\frac{GM}{rc^2}\label{eq:taylor}
\end{equation}

More precisely, the predictions of the Einstein-Hilbert equation in the solar system
are often tested in the PPN framework or a Yukawa fifth force
framework. The simplest PPN framework is characterized by two constant
parameters $\beta$ and $\gamma$ inserted in the previous Taylor expansion
(\ref{eq:taylor})
\begin{equation}
  \label{eq:5}
  g_{00}=1+2\phi+2\beta\phi^2+\ldots\quad,\quad-g_{rr}=1-2\gamma\phi+\ldots
\end{equation}

\noindent It turns out that 30 years of tests have put stringent
bounds on the parameters $\beta$ and $\gamma$ and selected a vicinity
of GR when analyzed in the PPN framework, namely
$\gamma-1=(2.1\pm2.3)\times10^{-5}$ and
$\beta-1=(-2.5\pm7.5)\times10^{-5}$.  Concerning $\gamma$, the current
bound is essentially given by the experiment performed through radar
ranging of the Cassini probe during its 2002 solar occultation [3],
while the bound on $\beta$ is obtained via analysis of
ephemerides [4].

It is important to note that the previous observations have also been
tested in other frameworks. For example the so-called ``fifth force''
tests which focus on a possible scale-dependent deviation from the
gravity force law. Such a deviation, corresponding to an additional
massive scalar gravity field, reduces to a modification of the
Newtonian potential by an additional Yukawa potential
\begin{equation}
  \label{eq:7}
  g_{00}=[g_{00}]_\GR+2\phi(r)\alpha
\exp\left(-\frac{r}{\lambda}\right)
\end{equation}

\noindent with an amplitude $\alpha$ measured with respect to Newton
potential and a range $\lambda$ related to the mass scale of the
hypothetical new particle which would mediate the ``fifth force''. The
presence of such a correction has been looked for on a large range of
distances and it turns out that the Yukawa term is excluded with a
high accuracy at some ranges, for example $\alpha<10^{-10}$ at
$\lambda\simeq$ Earth-Moon distance and $\alpha<10^{-9}$ at
$\lambda\simeq$ Sun-Mars distance. These bounds, again deduced from
Lunar Laser Ranging and tracking of planetary probes, correspond to a
remarkable result which approaches the accuracy of equivalence
principle tests.

\vspace*{0.7cm}
\noindent {\large 2. NEW FRAMEWORK OF METRIC EXTENSION OF GR}
\smallskip

We now present a more general framework, which extends the two
previously discussed frameworks. As already emphasized, the form of the coupling
between energy-momentum and curvature can still be discussed so that we can
generalize Einstein-Hilbert equation (\ref{eq:EI}) such that it takes
the form of a non local response relation between Einstein
curvature and the energy-momentum tensor (see~[5,6] for details)
\begin{equation}
  \label{eq:10}
  G_{\mu\nu}(x)=\int\dd^4x'\chi_{\mu\nu}^{\;\;\;\;\rho\sigma}(x-x')T_{\rho\sigma}(x')
\end{equation}

\noindent As an example, we retrieve GR with the following local
expression of the susceptibility $\chi_{\mu\nu}^{\;\;\;\;\rho\sigma}$
\begin{equation}
  \label{eq:11}
  \left[\chi_{\mu\nu}^{\;\;\;\;\rho\sigma}(x-x')\right]_\GR=\frac{4\pi G_N}{c^4}\left(
\delta_\mu^\rho\delta_\nu^\sigma+\delta_\nu^\rho\delta_\mu^\sigma\right)\delta^{(4)}(x-x')
\end{equation}

Let us note at this point that generalized response equations
naturally arise from radiative corrections of GR [5,6].  Radiative
corrections may induce modifications of GR not only at high energies,
but also at large distances [7]. In our simple model of solar system,
the stress tensor reads $T_{\rho\sigma}(x)=\delta_{\rho0}
\delta_{\sigma0}Mc^2\delta^{(3)}(\bm{x})$ so that the modified
Einstein tensor reads
\begin{equation}
  \label{eq:13}
  \delta G_{\mu\nu}(x)=Mc^2\int c\dd
  t'\;\delta\chi_{\mu\nu}^{\;\;\;\;00}(x-x')\quad,\quad x'=(ct',\bm{0})
\end{equation}

An important feature of this framework is that the Einstein tensor
$G_{\mu\nu}$, and therefore the Ricci curvature $R_{\mu\nu}$ do no
longer vanish outside the sources, as it is the case in
GR. Neverthelesse, because GR is in good accordance with observations,
it is expected that $\delta G_{\mu\nu}$ should stay small. Within a
linear approximation, the Bianchi identity can be used to extract the
two degrees of freedom $\delta G^{(0)}$ and $\delta G^{(1)}$ which are
present~:
\begin{equation}
  \label{eq:14}
  \delta G_{\mu\nu}(r)=\delta G_{\mu\nu}^{(0)}(r)+\delta G_{\mu\nu}^{(1)}(r)
  =\pi^{(0)}_{\mu\nu00}\frac{8\pi M\delta G^{(0)}(r)}{c^2}+
  \pi^{(1)}_{\mu\nu00}
\frac{8\pi M\delta G^{(1)}(r)}{c^2}
\end{equation}

\noindent where $\pi^{(0)}_{\mu\nu00}$ and $\pi^{(1)}_{\mu\nu00}$ are
operators of projection on the two sectors of different conformal weights, that is
to say the traceless and traced part (see [5] for their
expression). It has been shown in [6] how these two functions $\delta G^{(0)}$ and
$\delta G^{(1)}$ are related to a modification $\delta w$ of $w\equiv
\frac{1}{2}\ln\vert g_{rr}/g_{00}\vert$ (which is equivalent to a modification of the Weyl tensor), 
and a modification $\delta R$ of the Ricci scalar $R$
\begin{equation}
  \label{eq:18}
  \delta G^{(0)}(r)=-\frac{c^2}{8\pi
    M}\left(\partial_r^2+\frac{2}{r}\partial_r\right)\delta w(r)\qquad,\qquad
\delta G^{(1)}(r)=-\frac{c^2}{8\pi
    M}\delta R(r)
\end{equation}

Generally speaking, in the present framework, the gravitational
constant $G_N$ is replaced by a non local susceptibility tensor of
rank 4 which reduces to only two functions $\delta G^{(0)}$ and
$\delta G^{(1)}$. Those two functions correspond to two different
sectors for gravitation which, in the simplified situation considered
here, can be exactly matched to the two degree of freedom $g_{00}$ and
$g_{rr}$ with the help of (\ref{eq:18}).
\begin{equation}
  \label{eq:15}
  g_{00}=[g_{00}]_\GR+\delta g_{00}\qquad,\qquad g_{rr}=[g_{rr}]_\GR+\delta g_{rr}
\end{equation}

The functions $\delta g_{00}(r)$ and $\delta g_{rr}(r)$ can have
a general $r$-dependence. As a pedagological exemple, we will
focuss in the following on simplified phenomenological models obtained as a
Taylor expansion of $\delta g_{00}$ and $\delta g_{rr}$ in a vicinity
of the Solar system ($\alpha_nr^n\ll1$ and $\chi_nr^n\ll1$)
\begin{equation}
  \label{eq:16}
  \delta g_{00}=2\sum_{n>0}\alpha_n r^n\qquad,\qquad \delta g_{rr}=2\sum_{n>0}\chi_n r^n
\end{equation}

\noindent More elaborated models can be obtained by adding a cutoff to
allow the perturbation to start from a given distance from the
sun. The previous models can be viewed equivalently as distance-dependent PPN
parameters
\begin{equation}
  \label{eq:17}
  \beta(r)-1=\left(\frac{c^2}{GM}\right)^2\sum_{n>0}\alpha_n
  r^{n+2}\qquad,\qquad \gamma(r)-1=-\frac{c^2}{GM}\sum_{n>0}\chi_n
  r^{n+1}
\end{equation}

\vspace*{0.7cm}
\noindent {\large 3. PHENOMENOLOGICAL CONSEQUENCES ON A SIMPLIFIED MODEL}
\smallskip

As emphasized in the previous section, modifications of $g_{00}$ are heavily
constrained. Therefore, we will
consider in the following a modification of the spatial part
$g_{rr}$ alone. Then, even if we can deal with the general case, we
further restrict ourselves to the following simple modification
\begin{equation}
  \label{eq:1}
  \delta g_{rr}=2\chi_2r^2
\end{equation}

This specific model corresponds to a constant Ricci curvature
$R_{rr}=8\chi_2$, so that $\chi_2$ represents the amount of Ricci
curvature outside the source. The geodesic equation is modified, as
well as the Shapiro time delay, leading to anomalies in spacecraft
tracking and ephemerides. For example, a test particle (spacecraft or
planet) is submitted to the following anomalous coordinate
acceleration
\begin{equation}
\delta a_r=2\chi_2\left(rv_r^2-rv_\theta^2-GM\right) \quad,\quad
\delta a_\theta=4\chi_2\,r v_rv_\theta\label{eq:acc}
\end{equation}

\noindent It is easy to verify that if $\chi_2>0$, a purely radial
escape trajectory suffers a positive radial acceleration (towards the outside of the
solar system) while a circular orbit suffers a negative radial
acceleration (towards the sun). Moreover, the orthoradial acceleration
$\delta a_\theta$ vanishes for circular orbits and radial escape trajectories. The
order of magnitude of the anomalous acceleration is given by
\begin{equation}
  \label{eq:6}
  \delta a\sim\chi_2GM
\end{equation}

This anomalous coordinate acceleration is of course not the observable
that is deduced from measurement. Indeed, one has to take into account
the effect on light propagation. The modification $\delta \mathcal{T}$ of the one-way
light-time $\mathcal{T}=t_2-t_1$ from position $\bm{r}_1(t_1)$ to
position $\bm{r}_2(t_2)$ can be computed by integrating $\dd s^2=0$
along the path, giving
 \begin{equation}
  c\mathcal{T}=R_{12}+c\delta\mathcal{T}_{\text{Shapiro}}+c\delta\mathcal{T}
  \quad,\quad c\delta\mathcal{T}=-\frac{1}{3}\,\chi_2
  R_{12}\left(r_1^2+r_2^2+\bm{r}_1\cdot\bm{r}_2\right)
\end{equation}

\noindent where $r_i=\vert\bm{r}_i\vert=[r_i]_\GR+\delta r_i$ and
$R_{12}=\vert \bm{r}_2-\bm{r}_1\vert=[R_{12}]_\GR+\delta R_{12}$ are
modified when compared with their GR values due to the anomalous
coordinate acceleration (\ref{eq:acc}). The tracking observables can
then be computed from the knowledge of the two-way light-time (a sum
of two one-way light-time connected at the level of the
spacecraft). For example, up to small relativistic corrections, the
Doppler signal is obtained by the time derivative of the light-time while the
position in the sky (right ascension and declination) is obtained
through spatial derivatives.

\begin{figure}[h]
\begin{center}
\includegraphics[scale=0.8]{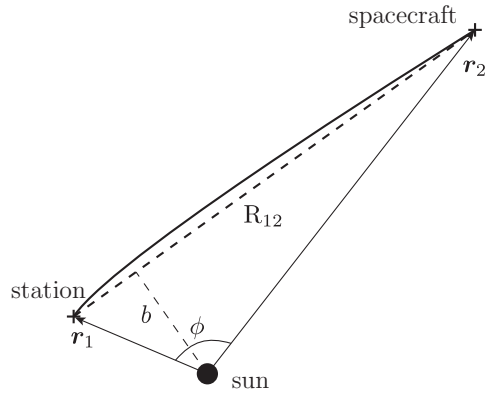}
\caption{Geometry of spacecraft tracking.}
\end{center}
\end{figure}

An important feature is that the light-time correction naturally
produces daily and semi-annual modulations, due to the geometry of the
triangle sun-station-spacecraft. Those modulations will show up in all
observables, in particular in ranging. Therefore, correlated anomalies
should be observed in ephemerides and analyzed within this
framework. As a matter of fact, only a careful comparison of
observations in the outer solar system, within the post-Einsteinian
phenomenological framework, could allow one to determine whether all
gravity tests can be compatible with the anomalies seen in the solar
system. This comparison will have to account for the presence of the
two sectors as well as for their $r$-dependences.



\vspace*{0.7cm}

\noindent {\large 4. REFERENCES}

%
%
%
%
%
{

\leftskip=5mm
\parindent=-5mm

\smallskip

[1] J.G. Williams, X.X. Newhall and J.O. Dickey, Phys. Rev. D 53, 6730
(1996).

[2] R.W. Hellings et al, Phys. Rev. Lett. 51, 1609 (1983); J. D. Anderson et al , Astrophys. J. 459, 365 (1996).

[3] Bertotti B., Iess L. and Tortora P., Nature 425 374 (2003).

[4] Fienga A. et al, published in this proceeding; Folkner W.-M, IAU
symposium, volume 261, 155 (2010); Pitjeva E., IAU symposium, volume
261, 170 (2010)

[5] M.-T. Jaekel and S. Reynaud, Annalen der Physik 4, 68 (1995).

[6] M.-T. Jaekel and S. Reynaud, Class. Quantum Grav. 23, 777 (2006); Class. Quantum Grav. 22 (2005) 2135-2157.

[7] G. t'Hooft and M. Veltman, Ann. Inst. H. Poincar\'e A 20, 69
(1974); E.S. Fradkin and A.A. Tseytlin, Nucl. Phys. B 201, 469 
(1982); O. Lauscher and M. Reuter, Class. Quantum Grav. 19, 483
(2002); H.W. Hamber and R.M. Williams, Phys. Rev. D 75, 084014 (2007).

}

\end{document}